ARTICLE                                                                                        Open Access

# Acoustic radiation-free surface phononic crystal resonator for in-liquid low-noise gravimetric detection

Feng Gao[1], Amine Bermak[1], Sarah Benchabane[2], Laurent Robert[2] and Abdelkrim Khelif[2]

## Abstract

Acoustic wave resonators are promising candidates for gravimetric biosensing. However, they generally suffer from strong acoustic radiation in liquid, which limits their quality factor and increases their frequency noise. This article presents an acoustic radiation-free gravimetric biosensor based on a locally resonant surface phononic crystal (SPC) consisting of periodic high aspect ratio electrodes to address the above issue. The acoustic wave generated in the SPC is slower than the sound wave in water, hence it prevents acoustic propagation in the fluid and results in energy confinement near the electrode surface. This energy confinement results in a significant quality factor improvement and reduces frequency noise. The proposed SPC resonator is numerically studied by finite element analysis and experimentally implemented by an electroplating-based fabrication process. Experimental results show that the SPC resonator exhibits an in-liquid quality factor 15 times higher than a conventional Rayleigh wave resonator at a similar operating frequency. The proposed radiation suppression method using SPC can also be applied in other types of acoustic wave resonators. Thus, this method can serve as a general technique for boosting the in-liquid quality factor and sensing performance of many acoustic biosensors.

## Introduction

The rapid and decentralized detection of biomolecules has been increasingly demanded for various applications, such as infectious disease diagnosis and food safety tests. This demand has been particularly evident during the recent outbreak of the novel coronavirus (COVID-19), where the throughput of time-consuming laboratory virus tests significantly delayed the diagnosis of the disease. In recent years, various techniques have been developed to meet these rising needs, which can be classified into four major categories: electrochemical, thermal, optical, and mass-sensitive biosensors[1,2]. Electrochemical biosensors detect signal variations in potential, current, or conductivity[3]. Thermal biosensors use the biochemical-reaction-induced temperature variation as the signal[4]. Optical biosensors detect the adsorption or emission of light at specific wavelengths[5]. Mass-sensitive biosensors sense surface mass variation due to the binding of analytes to bioreceptors[6]. Novel types of biosensors, such as photothermal[7] and photoelectrochemical[8–10] biosensors, have also been increasingly studied. These sensors use light as the excitation source while implementing detection based on temperature (photothermal) or electrical current (photoelectrochemical) variations. Among these different types, optical biosensors based on fluorescent labels or the label-free surface plasmon resonance technique are the most widely used for protein and aptamer detection[11–13]. However, due to the relatively complex setup of the optical detection scheme, optical sensors remain relatively costly and hard to miniaturize.

Mass-sensitive biosensors based on acoustic wave resonators are competitive alternatives to optical biosensors[14,15]. They can reach a higher level of integration at lower cost, as they do not require the use of peripheral equipment, such as

Correspondence: Feng Gao (fgao@hbku.edu.qa)
[1]College of Science and Engineering, Hamad Bin Khalifa University, Education City, Doha, Qatar
[2]Institut FEMTO-ST, CNRS, Université de Bourgogne-Franche-Comté, Besançon, France

excitation light sources. As their operating principle relies on the direct detection of an added mass to the surface of the sensor and on the direct measurement of the corresponding electroacoustic response, the entire detection scheme is intrinsically embedded into the acoustic device, leading to small-sized, low-cost sensors that can be easily fitted into a small microfluidic chamber. In addition, as mass variation is a physical signal that exists in any type of analyte–bioreceptor binding reaction, mass-sensitive sensors can be applied to all types of biomolecule detection, while other biosensors are limited by their signal type and require the specific design of the detection protocol. For example, potentiometric electrochemical biosensors require the analyte binding reaction to generate a potential variation so that it can be detected. Despite these advantages, acoustic wave resonators generally suffer from strong acoustic radiation in liquid, which decreases their quality factor (Q) and increases the signal noise. To maintain operation in liquid, acoustic wave resonators for biosensing are usually designed to operate in shear modes. When the movement of the shear wave is parallel to the liquid–solid interface, the mechanical motion transferred to the liquid is reduced compared to vertically polarized waves[16–18]. The acoustic radiation, however, cannot be eliminated, as the horizontal friction between the solid and water particles still conveys energy. The complete suppression of acoustic radiation can only be achieved by completely preventing wave propagation in the liquid. This scenario requires the velocity of the acoustic wave generated in the solid substrate to be lower than the sound velocity in water, which does not occur in natural materials. The acoustic wave velocity in all piezoelectric substrates is indeed always higher than the sound velocity in water because of the large elastic constants of solid materials[19].

It was, however, reported that one-dimensional surface phononic crystals (SPCs) made of periodic high aspect ratio electrode strips can be used to significantly slow down Rayleigh waves[20] and Lamb waves[21]. In this paper, we exploit this idea for the realization of SPC resonators operating in liquid. The proposed SPC resonator is theoretically studied by the finite element method (FEM) and experimentally implemented by the classical lithography, electroplating, and molding (Lithographie, Galvanoformung, Abformung (LIGA)) process. By incorporating SPC with interdigitated transducers (IDTs), we find that the velocity of Rayleigh waves can be reduced to a value lower than the velocity of the sound in water. This successfully stops the propagation of the acoustic wave in water and eliminates acoustic radiation. Because of the complete suppression of radiation, the in-liquid Q factor of the resonator is improved by more than 15 times compared to a conventional Rayleigh wave resonator working in the same frequency range. In addition to the slowing down of the phase velocity, the group velocity is also found to be reduced to almost zero, hence suppressing energy propagation in the horizontal plane. This result enables the use of zero or a small number of reflectors when constructing a resonator, therefore significantly reducing the sensor size and fabrication cost. Moreover, the high aspect ratio electrodes constituting the SPC naturally lead to an increase in the surface-to-volume ratio of the device and hence increase the mass sensitivity. The proposed acoustic radiation suppression method can also be applied in other types of acoustic waves, which makes it a general technique that can significantly push forward the performance limit of many acoustic biosensors.

## Results
### Design of the SPC resonator

Figure 1a shows a top-view diagram of the SPC resonator. IDTs consisting of 30 pairs of high aspect ratio electrodes located in the center of the device are deposited on a 128° Y-cut lithium niobate substrate. Nickel is chosen as the electrode material because high aspect ratio nickel electrodes can be reliably implemented using the LIGA process. When an electrical field is applied to the IDTs, mechanical deformations are stimulated in the piezoelectric substrate, eventually forming phononic resonance. Both the electrode width and the electrode spacing are equal to 2.5 μm, resulting in a wavelength (λ) of 10 μm. The electrode height ($t_{ele}$) is set to ~7.5 μm. This value is much larger than the thickness of conventional SAW resonators, which are usually one or two percent of the wavelength[22]. Because of the low group velocity in the horizontal plane, only 20 reflector strips are placed on the two sides of the IDTs to enhance the resonance. This number is much smaller than that used in conventional SAW resonators, which usually require hundreds of reflector strips[23]. The resonance frequency of the device is determined together by the electrode geometry, electrode periodicity, and materials of the electrode and piezoelectric substrate.

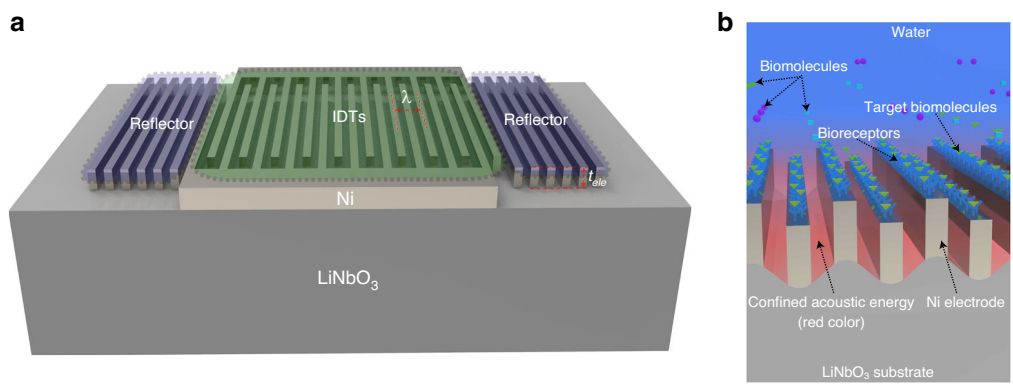

**Fig. 1 Diagram and sensing mechanism of the SPC resonator. a** Diagram of the SPC resonator. IDTs for wave stimulation are in the center, while reflectors for resonance enhancement are on the two sides. **b** Diagram of the SPC used for biomolecule detection. The specific binding between bioreceptors and target biomolecules induces mass loading to the device surface, which causes a resonance frequency shift.

Figure 1b shows a diagram illustrating how the SPC resonator can be used for biosensing. To detect biomolecules, bioreceptors are immobilized on the surface of the electrodes. These bioreceptors can bind specifically with their detection targets. Typical bioreceptors are DNA probes, antigens, and antibodies[1,24]. They are widely used in the detection of human immunoglobulins or pathogens, such as viruses and bacteria. Less specific bioreceptors, such as functionalized nanoparticles[25] and bioimprinted polymers[26,27], can be used for the detection of small biomolecules that can serve as biomarkers for early disease diagnosis. Once the detection targets bind with the bioreceptors, the additional mass attached to the device surface results in a twofold variation in the SPC resonance. First, the attached molecules increase the mass of the high aspect ratio electrodes, which reduces their mechanical resonance frequency. Second, the additional mass reduces the surface acoustic wave velocity due to the classical mass loading effect also observed in surface acoustic wave resonators[18]. As the SPC resonance is a coupling between the mechanical resonance of the high aspect ratio electrodes and the surface acoustic wave, the combination of the above two effects reduces the SPC resonance frequency. By building an oscillator-based frequency readout circuit with the SPC resonator[28,29], the sensor response can be converted to real-time digital signals.

### Slow acoustic wave in the SPC

If the acoustic wave velocity is reduced to a value lower than the sound velocity in water, its propagation in water is inhibited. This eliminates the acoustic radiation of the wave and thus enables high Q resonance in liquid. By exploiting this principle, we used SPC made of periodic high aspect ratio electrodes to slow down the surface wave on a lithium niobate substrate. The SPC induces a hybridization between the Rayleigh-type surface wave and the elastic resonance of the high aspect ratio electrodes. This hybridization results in the occurrence of hybrid modes with velocities directly conditioned by the geometrical characteristics of the electrodes. Figure 2a shows the simulated resonance frequency of the localized mode of the SPC resonator in air and water for different electrode heights. It can be seen from the triangular-marked blue curve that the resonance frequency decreases continually with an increasing electrode height in air. However, this mode cannot be observed in water for electrode heights lower than 5.6 μm, for which the acoustic wave velocity ($v = \lambda \times f$) on the substrate remains higher than the velocity in the liquid. For electrode heights larger than 5.6 μm, in-liquid resonance occurs, as seen from the circle-marked orange curve. This result is because the corresponding wave velocity in the piezoelectric substrate is lower than the sound velocity in water, which prevents acoustic radiation in the liquid. The corresponding displacement fields in the substrate and acoustic pressure distribution in the liquid were then simulated. The results obtained for electrode heights of 2 and 7.6 are reported in Fig. 2b and c, respectively. In the case of the 2-μm electrodes, acoustic wave generation and radiation were simulated at 270 MHz. As shown by the alternating positive (red part) and negative (blue part) acoustic pressures, the acoustic wave propagates in water until it reaches the perfectly matched layer (PML) boundary, where the wave is absorbed by the PML. It should be noted that 270 MHz is the resonance frequency of the 2-μm electrode device in air. This frequency is chosen because the device cannot resonate in water. In comparison, the acoustic pressure distribution of the SPC resonator with the 7.6-μm electrode operating in water at its resonating frequency (124.2 MHz) is shown in Fig. 2c. It can be seen that the positive (red part) and negative (blue part) acoustic pressures exist only in the region near the electrodes, which means that the acoustic energy is confined in this region and the acoustic wave is not radiative. Figure 2c also reveals the mode shape of the SPC resonators. The displacements of the electrodes are much larger than those observed at the piezoelectric substrate surface, which means that most of the elastic energy is stored in the electrodes. The transition from surface-confined energy to electrode-surface hybridized energy reveals how the wave is transformed from a conventional Rayleigh SAW into an interfacial wave, where the elastic energy is distributed between the high aspect ratio electrode and the near surface of the substrate. The movement of the electrodes is vertically polarized, as they move up and down in an oscillation cycle. The displacements of the electrodes in a full cycle are shown in Fig. S1, Supplementary Information. Although the vertical movement transfers more mechanical energy to water compared to the horizontal movement in shear waves, resonance in water still occurs. This behavior indirectly suggests the effectiveness of the acoustic radiation suppression technique.

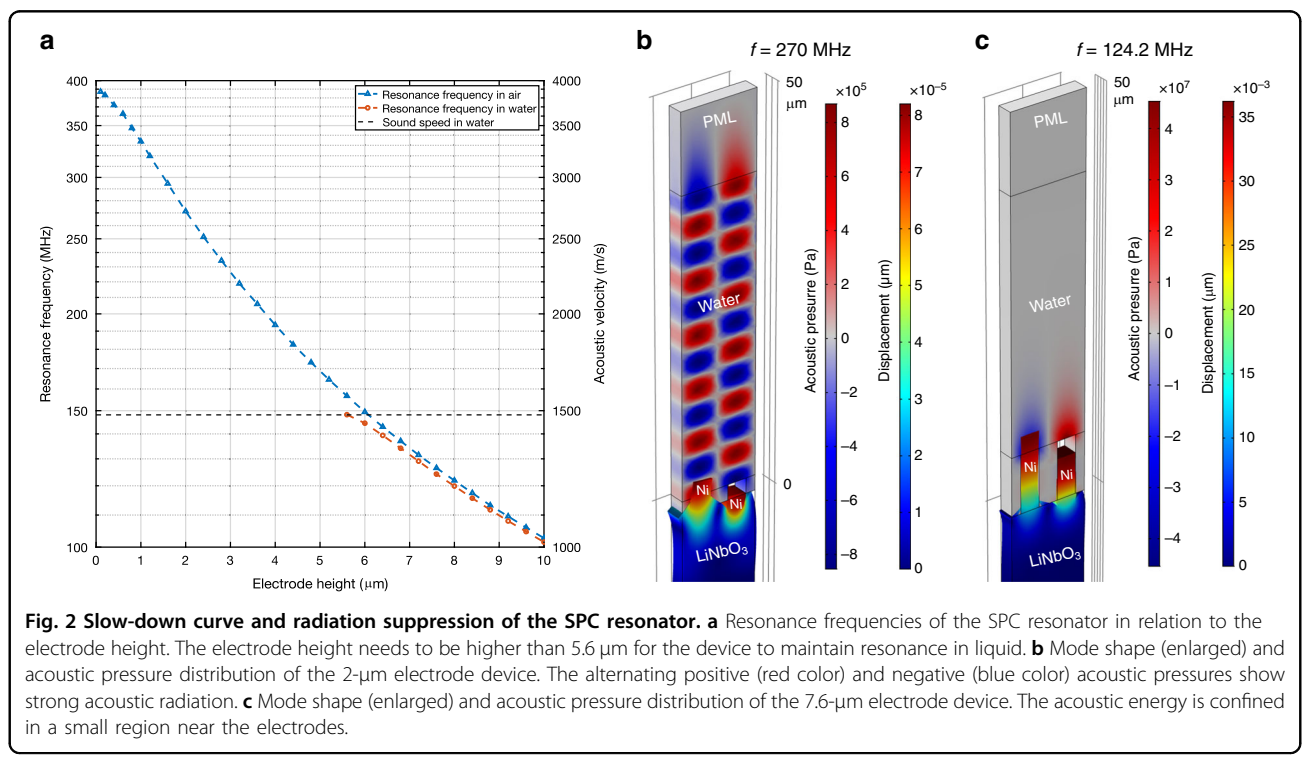

**Fig. 2 Slow-down curve and radiation suppression of the SPC resonator. a** Resonance frequencies of the SPC resonator in relation to the electrode height. The electrode height needs to be higher than 5.6 μm for the device to maintain resonance in liquid. **b** Mode shape (enlarged) and acoustic pressure distribution of the 2-μm electrode device. The alternating positive (red color) and negative (blue color) acoustic pressures show strong acoustic radiation. **c** Mode shape (enlarged) and acoustic pressure distribution of the 7.6-μm electrode device. The acoustic energy is confined in a small region near the electrodes.

Dispersion curves providing information on the group and the phase velocities constitute a very important design tool for acoustic devices. Figure 3a shows two individual impedance curves of an SPC resonator with a 7.6-μm thick electrode when $k_x$ equals zero and $0.2\pi/\lambda$. By varying the wavevector $k_x$ in the first Brillouin zone, the dispersion curves (Fig. 3b) of these resonance modes are obtained. As an example, the two impedance curves in Fig. 3a yield the dispersion curve points marked by two vertical dashed black lines in Fig. 3b. Because of the electrical periodicity of the electrodes, the dispersion curves fold at $k_x = \pi/\lambda$, which means that its folded upper strand is the dispersion curve of $\pi/\lambda < k_x < 2\pi/\lambda$. By checking the corresponding mode shape, the red-dashed line in Fig. 3b is found to be the dispersion curve of the vertically polarized wave, while the yellow-dashed line is the dispersion curve of the shear horizontal wave. The red dot in Fig. 3b thus marks the base mode of the vertical polarized phononic wave, where $k_x = 2\pi/\lambda$. It can be seen from the slope of the dispersion curve that the corresponding group velocity of this point is almost zero (slightly negative). As the group velocity represents the transmission of the wave energy, this significantly reduced group velocity along the $x$ direction means that very little energy propagates horizontally. Because of this, the use of a large number of reflectors, as in conventional SAW resonators, is not necessary. This result enables a significant reduction in the device size and saves cost.

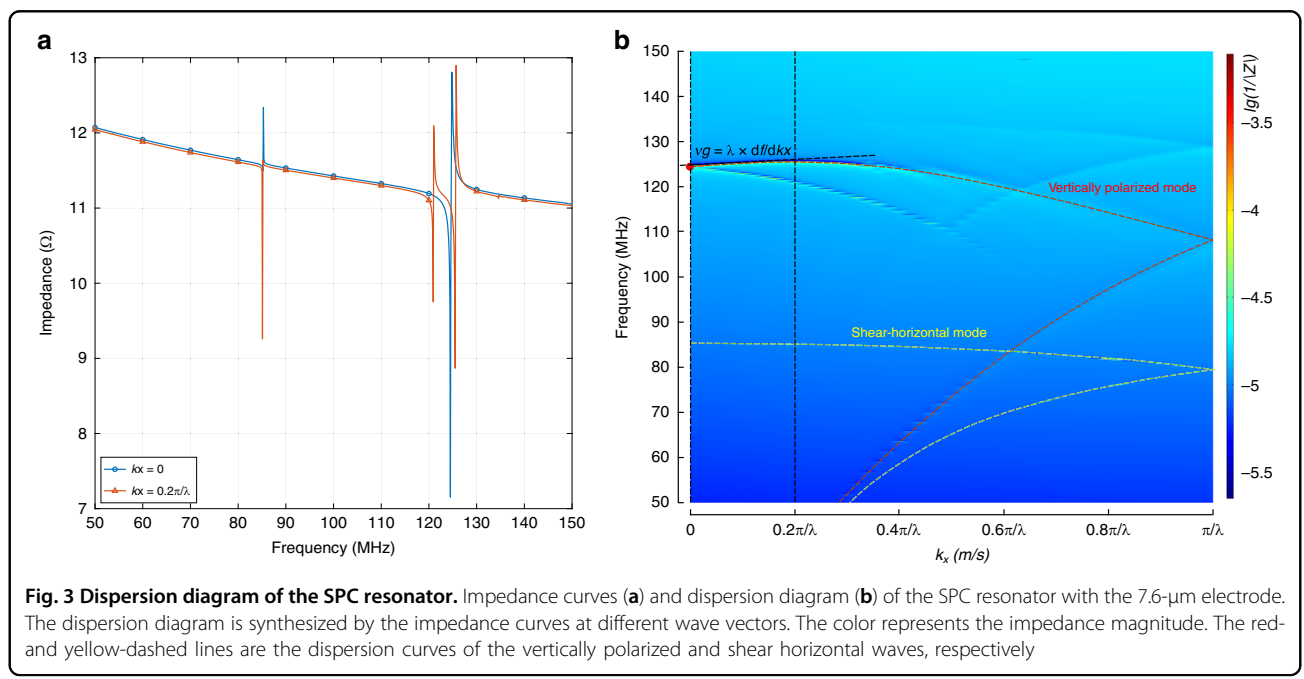

**Fig. 3 Dispersion diagram of the SPC resonator.** Impedance curves (**a**) and dispersion diagram (**b**) of the SPC resonator with the 7.6-μm electrode. The dispersion diagram is synthesized by the impedance curves at different wave vectors. The color represents the impedance magnitude. The red- and yellow-dashed lines are the dispersion curves of the vertically polarized and shear horizontal waves, respectively

### Electrode profile and impedance of the SPC resonator

A scanning electron microscopy (SEM) image of the fabricated device is shown in Fig. 4a. Both positive photoresist (pPR) and negative photoresist (nPR) were used as electroplating molds to produce different electrode profiles and investigate the corresponding impact on the resonator performance. The cross-sectional SEM images of the electrodes fabricated with the pPR and nPR are shown in Fig. 4b and c, respectively. The cross-section was obtained by cutting with a focused ion beam, during which platinum was deposited to protect the adjacent top surface of the electrode. The platinum protection layer and the nickel electrode are marked by blue and red overlays in one of the electrodes of each image, respectively. The profile of the electrode fabricated by nPR is a trapezoid with a slightly shrunken bottom. In comparison, a reverted trapezoidal profile is obtained with pPR.

The impedance curves of the resonators fabricated with these different photoresists are shown in Fig. 4d. The two SPC resonators have similar impedance characteristics. The difference in resonance frequency is caused by both the electrode profile and electrode thickness differences of the devices. For comparison, the response of a conventional Rayleigh wave resonator is also shown. This resonator is fabricated on the same 128° Y-cut lithium niobate wafer with a one-step lift-off process. The wavelength and electrode thickness are 30 μm and 200 nm, respectively. Its layout design is directly scaled up three times from the SPC resonator for a fair comparison. The impedance curve of the conventional Rayleigh wave resonator is much flatter than that of the SPC resonators, with a Q factor in liquid that only reaches 3.4 due to its acoustic radiation in liquid. In contrast, the Q factors of the SPC resonators in liquid are 50.4 and 44.5 for the pPR and nPR processes, respectively. The sharp resonance peaks mean that the SPC resonators have better frequency resolution while detecting the external mass loads. In practical applications, oscillators incorporating sensor resonators are usually built to produce frequency signals that can be counted by readout circuits. The phase noise of the oscillator is inversely proportional to the Q factor of the resonator and can be estimated by Leeson's equation[30]. Resonators with high quality factors will lead to lower frequency noise in the final sensor oscillator.

### Mass sensitivity of the SPC resonator

The overall performance of a gravimetric biosensor can be described by its limit of detection (LOD), which is obtained by

$$\text{LOD} = \frac{S}{n_f} \quad (1)$$

where $S$ and $n_f$ are the mass sensitivity and frequency noise of the sensor, respectively. The frequency noise of the sensor is evaluated by its Q factor, as discussed in the previous section. To evaluate the overall sensing performance, the mass sensitivity of the SPC biosensor also needs to be characterized. The variation in the nickel electrode height across the wafer is used here to calculate the mass sensitivity, as a change in electrode height is equivalent to the variation in loaded mass on top of the electrodes. Nevertheless, it should be noted that this sensitivity is an underestimate of the actual mass sensitivity expected from an actual biosensing process where the mass loading would be homogeneous on the entire device surface rather than localized on the top of the electrode. The resonance frequencies of the SPC resonators operating in air and in water are shown in Fig. 5a and b, respectively. The measurement data correspond to the triangular data points in the two figures. There are more data points of the resonators fabricated with pPR than for the devices fabricated with nPR because more units of the corresponding layout cells were arranged in the positive fabrication mask than in the negative fabrication mask. Linear regression was used to find the relations between the electrode height and resonance frequency, which are shown by solid lines in the two figures. Using the actual electrode profile shown in Fig. 4b, c, the resonance frequencies of the devices were also simulated to validate the effectiveness of the FEM model. The corresponding simulated resonance frequencies in air and in water are shown by the dashed lines in Fig. 5a and b, respectively. Both of the simulated resonance frequencies agree well with the experimental data. The mass sensitivity of the resonators to the loading on the top of the electrode is obtained from the slope of the fitted curve by the following equation:

$$S = \frac{1}{\rho_{Ni}} \times \frac{df}{dt} \quad (2)$$

where $\rho_{Ni}$ and $t$ are the density and height of the nickel electrode, respectively. The mass sensitivity is a negative number, as the mass loading results in a decrease in the

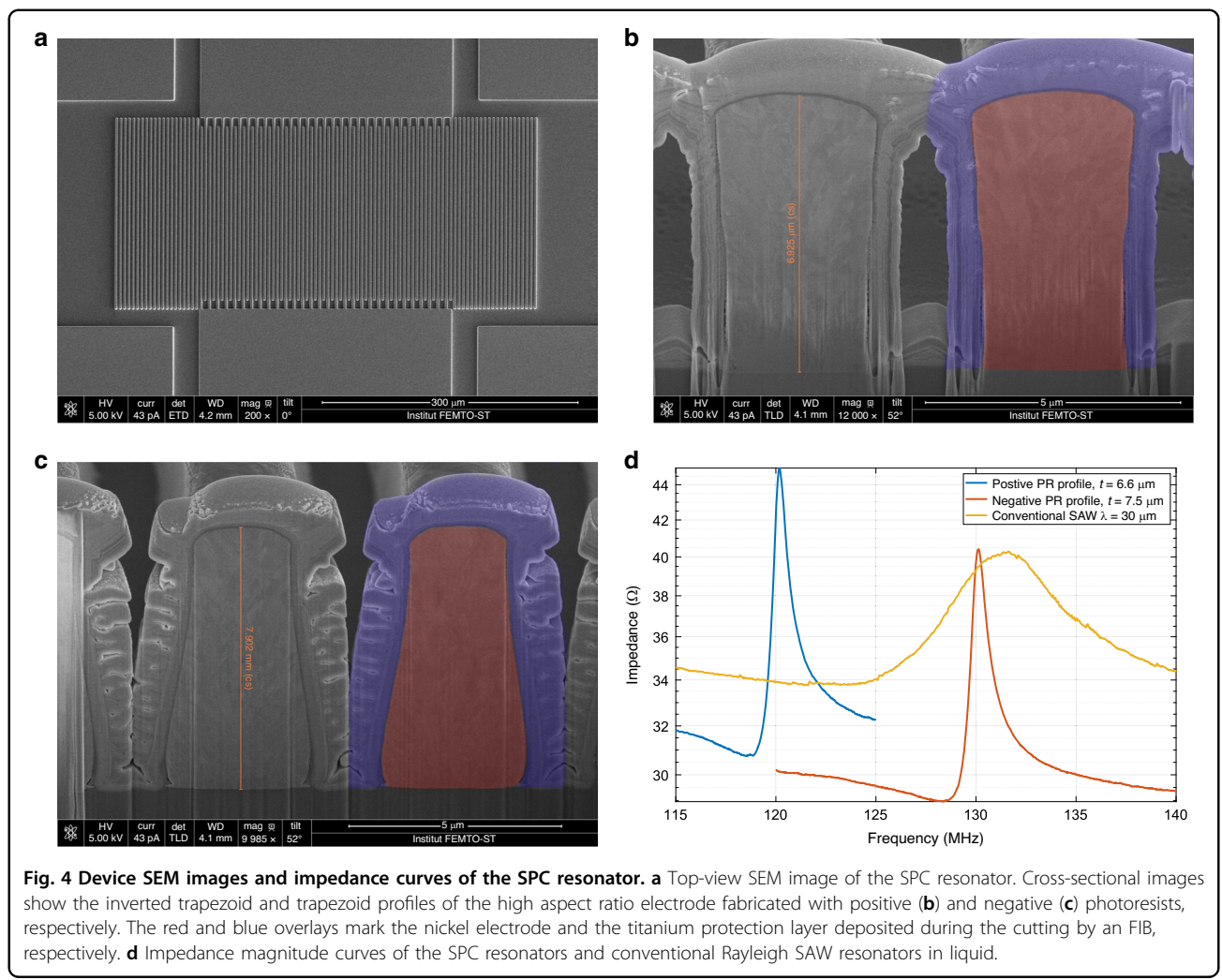

**Fig. 4 Device SEM images and impedance curves of the SPC resonator. a** Top-view SEM image of the SPC resonator. Cross-sectional images show the inverted trapezoid and trapezoid profiles of the high aspect ratio electrode fabricated with positive (**b**) and negative (**c**) photoresists, respectively. The red and blue overlays mark the nickel electrode and the titanium protection layer deposited during the cutting by an FIB, respectively. **d** Impedance magnitude curves of the SPC resonators and conventional Rayleigh SAW resonators in liquid.

resonance frequency. The mass sensitivities of the resonators in air ($S_a$) and in water ($S_w$) obtained from the linearly fitted curve are shown in Table 1. Their corresponding coefficients of determination ($R^2$) are also listed. All the $R^2$ values are greater than 0.94, which indicates a good match between the fitted line and experimental data. The data in Table 1 also reveal that the pPR SPC resonators have a mass sensitivity 30% higher than the nPR SPC resonators, whether operating in air or water. This result is mainly because the trapezoidal profile of the nPR process makes the mass-loaded area (the top surface) smaller than that obtained from the inverted trapezoidal profile of the pPR process. The mass sensitivity of the resonator in air is only slightly higher than the mass sensitivity of the resonator in water, which means that the water loading has a negligible impact on the mass sensitivity of the resonator. The simulated mass sensitivities in air ($S_{a\_sim}$) and in water ($S_{w\_sim}$) are also listed in Table 1. The sensitivity modeling for the pPR

SPC resonator agrees well with the experimental results, with discrepancies of 6.2 and 7.6% in air and water, respectively. The modeling for the nPR SPC resonators is less accurate, with simulated experiment discrepancies of 26.7 and 21.7% in air and water, respectively. The less accurate modeling of the nPR SPC resonator is possibly due to the discrepancy between the more distorted experimental electrode profile (Fig. 4c) and the simplified trapezoidal electrode profile in the simulation. Nevertheless, the FEM modeling results generally agree well with the experimental data.

The mass sensitivity under homogeneous loading conditions, or full coverage mass sensitivity, was then obtained by the same FEM simulation model. A polymer layer (PMMA) covering the entire device surface was added as the mass loading layer. The device frequency response as a function of added mass is simulated by changing the PMMA layer density. The simulated unit cells for the pPR and nPR SPC resonators operating at

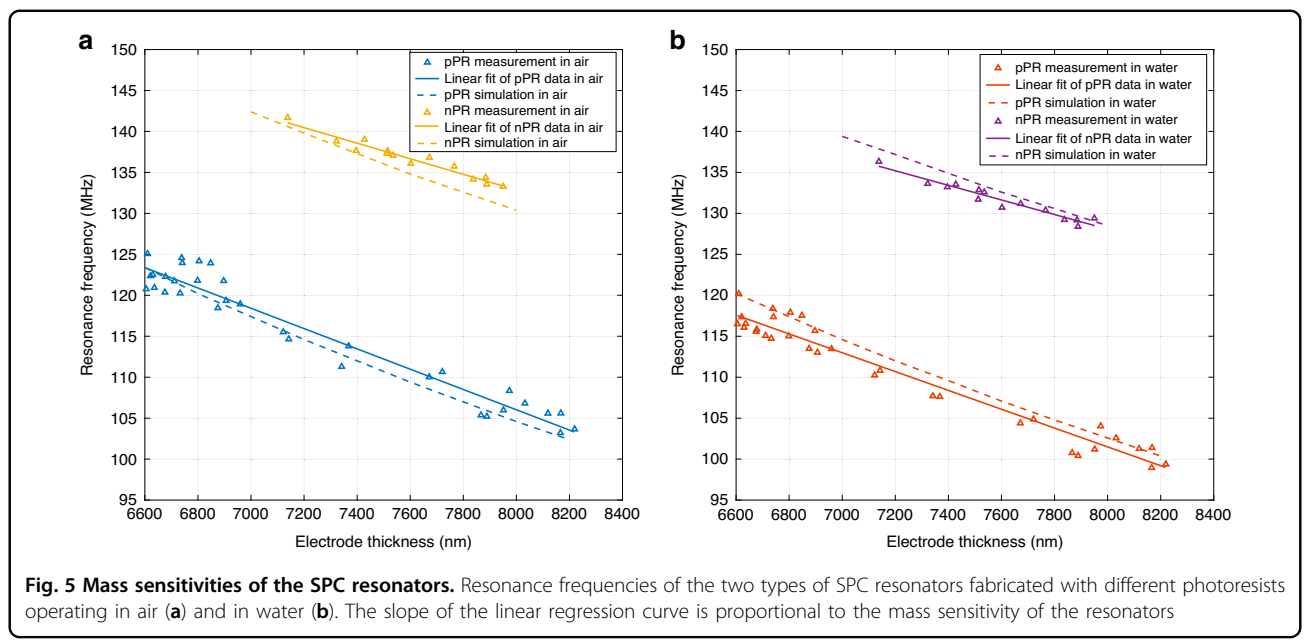

**Fig. 5 Mass sensitivities of the SPC resonators.** Resonance frequencies of the two types of SPC resonators fabricated with different photoresists operating in air (**a**) and in water (**b**). The slope of the linear regression curve is proportional to the mass sensitivity of the resonators

**Table 1** Mass sensitivity of the SPC resonators.

| Profile | $S_a$[a] | $R^2$ of $S_a$ | $S_w$ | $R^2$ of $S_w$ | $S_{a\_sim}$ | $S_{w\_sim}$ | $S_{w\_full}$ |
|---------|----------|----------------|-------|----------------|--------------|--------------|---------------|
| pPR | −13.89 | 0.94 | −12.92 | 0.95 | −14.75 | −13.90 | −56.7 |
| nPR | −10.64 | 0.94 | −10.07 | 0.94 | −13.48 | −12.25 | −58.4 |

[a]Unit of mass sensitivity: Hz/(ng/cm$^2$).

120 MHz are shown in Fig. S3a, b, respectively. As listed in the last column of Table 1, the simulated full coverage mass sensitivity in water ($S_{w\_full}$) is found to be −56.7 and −58.4 Hz/(ng/cm$^2$) for the pPR and nPR SPC resonators, respectively. The good agreement between the full coverage mass sensitivities of the pPR and nPR SPC resonators can be accounted for by their similar surface areas under full coverage conditions. This result also reveals that the electrode profile does not affect the mass sensitivity of the SPC when the mass loading is homogenous on the entire surface. As expected, the mass sensitivity for full surface coverage loading is significantly higher than the mass sensitivity for loading only on the top of the electrode because of the much larger loading area. For comparison, the mass sensitivity of the widely used 36° Y LiTaO$_3$-based SH-SAW sensing device, here working at 120 MHz, is also simulated. The corresponding simulation unit cell is shown in Fig. S3c. The result shows a mass sensitivity of 10.02 Hz/(ng/cm$^2$), which agrees well with the experimental results reported by Barie et al.[31]. Their 36° Y LiTaO$_3$ SH-SAW operating at 380 MHz showed a 25 kHz response to the adsorption of a bovine serum albumin monolayer that has a surface mass of 200 ng/cm$^2$. Considering that the mass sensitivity is proportional to the square of the operating frequency, the equivalent mass sensitivity at 120 MHz is 12.46 Hz/(ng/cm$^2$), which is similar to our simulated SH-SAW sensitivity. This comparison shows that the proposed SPC biosensor can exhibit a mass sensitivity close to six times higher than that of conventional SH-SAW biosensors, which is due to the higher surface-to-volume ratio of the high aspect ratio electrodes.

## Discussion

As mentioned in the "Introduction" section, the usual strategy for in-liquid acoustic sensing relies on the use of shear-polarized acoustic modes. The shear polarization indeed minimizes mechanical motion transfer to water. However, it does not fully prevent acoustic radiation in water: the wave velocity in the solid substrate remains higher than the sound velocity in liquid and coupling with the vertically polarized mode generally occurs. This radiation loss is one of the major limitations of these devices in achieving a high Q factor for in-liquid sensing. In contrast, the SPC resonator proposed in this work relies on the complete suppression of acoustic radiation in water; this suppression is accomplished by a significant slowing down of acoustic wave propagation in the solid substrate due to the hybridization mechanism between localized modes in the electrodes and the surface wave. Due to the complete suppression of acoustic radiation, the energy leakage can be theoretically fully stopped, hence enabling high Q factor in-liquid sensing. In the present work, the Q factor of the SPC resonators is mostly limited by the electroplating process. The sidewall surface roughness and the overall process inhomogeneity result in structural defects breaking the periodicity of the IDTs, which limits further improvement in the quality factor. Nevertheless, this technological shortcoming can be overcome, notably by using alternative thick-film deposition techniques, such as thermal evaporation and chemical vapor deposition. Another limitation of the SPC biosensor compared to conventional SAW devices is that it involves a more complex fabrication process. Although it is a trade-off to achieve better performance, the added cost will not be significant in a complete sensing system that includes other components, such as microfluidic channels and readout circuits.

The proposed method for acoustic radiation suppression is applied here to devices initially operating on elliptically polarized Rayleigh waves. This method can be more generally transposed to any IDT-based acoustic wave resonator, including SH-SAW or Love-wave acoustic resonators. The combination of shear mode polarization and acoustic radiation suppression has the potential to significantly advance the performance boundary of acoustic wave biosensors.

## Conclusion

In summary, we proposed a gravimetric biosensor based on an SPC resonator that can achieve acoustic radiation-free operation in liquid. The SPC induced a hybridization between a Rayleigh-type surface wave and the elastic resonance of the high aspect ratio electrodes, which led to strong confinement of the elastic energy at the electrode–substrate interface. The acoustic wave velocity was then reduced to be lower than the speed of sound in water. This mechanism resulted in the complete suppression of acoustic radiation, leading to high Q resonance in liquid. This device principle was validated by FEM analysis and experimentally implemented through the fabrication of SPC resonators using the LIGA process. Q factors on the order of 50 and a mass sensitivity of 12.92 Hz/(ng/cm$^2$) were obtained by the pPR process. A comparison with a Rayleigh SAW resonator revealed that the SPC could improve the Q factor of the resonator by 15 times. The effect was demonstrated here for resonators operating on vertically polarized waves, and it could also be combined with existing shear mode acoustic wave resonators. This proposed biosensor has the potential to significantly advance the performance boundary of current acoustic wave biosensors.

## Materials and methods
### Numerical analysis

Numerical analysis based on the FEM was used to study the behavior of the SPC resonator. The analysis was performed using Comsol multiphysics software. The FEM model couples multiple physics, including solid mechanics, pressure acoustics, electrostatics, piezoelectric effects, and acoustic structure interactions. As a 3D numerical analysis is computationally heavy, a simplified unit cell of the resonator (Fig. S2, Supplementary Information) was adopted to reduce the computation time. The width of the unit cell was set to one wavelength with continuity-type periodic conditions for the mechanical, electrical, and acoustic components applied to the two sides. This setup was equivalent to continuously repeating the unit cell and thus representing an infinite number of IDT pairs. The actual simulation model was in 3D with a depth of 0.1λ. Continuity-type periodic conditions were also applied to the front and back of the model to equivalently extend the depth to infinity, which effectively made the acoustic aperture infinitely wide. The thickness of the substrate ($t_{sub}$) was set to 5λ, with the bottom 1λ set as the PML. The PML absorbed all the waves leaking into it, equivalently making the substrate infinitely thick. A fixed boundary condition was applied to the bottom of the substrate, which represented the fixation on the accommodating package. The depth of the water ($d_{water}$) was also set to 5λ, with the top 1λ set as the PML to equivalently extend the water depth to infinity. A free boundary condition was applied to the top of the water. The height and profile of the electrodes were used as sweeping parameters in the analysis to study the behavior of the SPC resonator. The ideal periodic conditions and PMLs ruled out the performance impact from parameters of the substrate thickness, acoustic aperture width, and IDT designs, which made the analysis results only relevant to the profile and dimensions of the high aspect ratio electrodes. An alternative electrical potential (1-V amplitude) was applied to the two metallic electrodes as electrical stimulation.

Fundamental field parameters, such as the electrical fields, particle displacements, and acoustic pressures, were obtained from numerical analysis. Other needed information, including electrical impedance and mode shape, was further derived from those field parameters. To obtain the dispersion diagram of localized modes in SPC, Floquet periodic conditions were applied to the left and right sides of the model so that impedance curves under different wave vectors ($k_x$) were computed. The variation in $k_x$ covered the first Brillouin zone, i.e., from 0 to $\pi/\lambda$. This equivalently changed the periodic conditions on the two sides from continuity type ($k_x = 0$) to antiperiodic type ($k_x = \pi/\lambda$). The final dispersion diagram was obtained by a color map representation of the impedance magnitude in the $k_x - f$ plane.

## Fabrication process

The device was fabricated on a 4-inch 128° Y-cut lithium niobate wafer. The wafer was initially cleaned with a piranha solution and deionized water to remove potential contaminants. Following cleaning, 20-nm titanium and 100-nm copper thin films were deposited on the wafer surface. The copper layer was used as a seed layer for subsequently electroplating nickel, while the titanium layer was used to promote adhesion between the copper layer and lithium niobate substrate. Subsequently, photolithography was used to create the mold for electroplating. A thick pPR, AZ9260, and a thick nPR, AZ nLOF 2070, were used to create different electrode profiles. Two photolithography masks with opposite polarity were designed correspondingly. After the mold was created, 7–8 μm-thick nickel layers were electroplated by an automatic electroplating machine (Technotrans Microform 100). It should be noted that both the IDTs and bus bars were electroplated, which meant that the IDTs were mechanically clamped by the bus bars in their electrical connection nodes. The device was then immersed in a Microposit 1165 remover to strip the photoresist mold. Reactive ion etching (RIE) was then used to remove the copper and titanium seed layers between the nickel electrodes to define the electrically isolated IDT fingers. Subsequently, 200 nm of silicon dioxide was deposited on the sensor surface using chemical vapor deposition. This layer was used as a passivation layer and could also provide an interface for bioreceptor immobilization. Another photolithography step was then performed to create the photoresist etching mask for the area excluding the pads that will be used for external electrical connection. The removal of silicon dioxide covering the contact pads was completed by another RIE process. The complete process flow is shown in Fig. S4.

## Sensor characterization

The variation in the nickel film thickness across the wafer was exploited for the calculation of the mass sensitivity of the sensor because the small height difference of the nickel electrodes across the substrate caused a natural mass loading variation for the sensor, which changed its resonance frequency. The thickness of the nickel layers was measured by a profilometer.

The Q factor of the SPC resonator was used to evaluate the frequency stability (frequency noise) of the sensor. It was calculated from its impedance by the following equation:

$$Q = \frac{f}{2} \times \frac{d\phi}{df} \quad (3)$$

where $f$ and $\phi$ are the operating frequency and phase of the electrical impedance of the device, respectively. The measurement of the impedance was completed by a vector network analyzer (VNA). Two RF probes were used to connect the device to the VNA. To obtain the performance data of the sensor in both air and water, its impedance was measured under both conditions.

### Acknowledgements
This work is funded by NPRP grant no. NPRP10-0201-170315 from the Qatar National Research Fund (a member of Qatar Foundation). This work is also supported by the EIPHI Graduate School (contract "ANR-17-EURE-0002") and the French RENATECH network with its FEMTO-ST technological facility. The findings herein reflect this work and are solely the responsibility of the authors.

### Author contributions
F.G., A.B., and A.K. proposed the idea. F.G. and S.B. designed the fabrication process. F.G. and L.R. fabricated the device. F.G. performed the device test. F.G., A.K., and S.B. drafted the manuscript.

### Conflict of interest
The authors declare that they have no conflict of interest.

**Supplementary information** accompanies this paper at https://doi.org/10.1038/s41378-020-00236-9.

Received: 7 July 2020 Revised: 17 November 2020 Accepted: 13 December 2020
Published online: 18 January 2021

### References
1. Gaudin, V. Advances in biosensor development for the screening of antibiotic residues in food products of animal origin–a comprehensive review. *Biosens. Bioelectron.* **90**, 363–377 (2017).
2. Metkar, S. K. & Girigoswami, K. Diagnostic biosensors in medicine–a review. *Biocatal. Agric. Biotechnol.* **17**, 271–283 (2019).
3. Khan, M., Hasan, M., Hossain, S., Ahommed, M. & Daizy, M. Ultrasensitive detection of pathogenic viruses with electrochemical biosensor: state of the art. *Biosens. Bioelectron.* **166**, 112431 (2020).
4. Casadio, S. et al. Development of a novel flexible polymer-based biosensor platform for the thermal detection of noradrenaline in aqueous solutions. *Chem. Eng. J.* **315**, 459–468 (2017).
5. Khansili, N., Rattu, G. & Krishna, P. M. Label-free optical biosensors for food and biological sensor applications. *Sens. Actuators B: Chem.* **265**, 35–49 (2018).
6. Zhang, Y., Luo, J., Flewitt, A. J., Cai, Z. & Zhao, X. Film bulk acoustic resonators (FBARs) as biosensors: a review. *Biosens. Bioelectron.* **116**, 1–15 (2018).
7. Huang, L., Chen, J., Yu, Z. & Tang, D. Self-powered temperature sensor with seebeck effect transduction for photothermal–thermoelectric coupled immunoassay. *Anal. Chem.* **92**, 2809–2814 (2020).
8. Lv, S., Zhang, K., Zhu, L. & Tang, D. ZIF-8-assisted NaYF4: Yb, Tm@ ZnO converter with exonuclease III-powered DNA walker for near-infrared light responsive biosensor. *Anal. Chem.* **92**, 1470–1476 (2019).
9. Zhou, Q. & Tang, D. Recent advances in photoelectrochemical biosensors for analysis of mycotoxins in food. *Trends Anal. Chem.* **124**, 115814 (2020).
10. Shu, J. & Tang, D. Recent advances in photoelectrochemical sensing: from engineered photoactive materials to sensing devices and detection modes. *Anal. Chem.* **92**, 363–377 (2019).
11. Fan, X. et al. Sensitive optical biosensors for unlabeled targets: a review. *Anal. Chim. Acta* **620**, 8–26 (2008).
12. Chen, T., Wang, X., Alizadeh, M. H. & Reinhard, B. M. Monitoring transient nanoparticle interactions with liposome-confined plasmonic transducers. *Microsyst. Nanoeng.* **3**, 16086 (2017).
13. Oh, S. Y. et al. Development of gold nanoparticle-aptamer-based LSPR sensing chips for the rapid detection of *Salmonella typhimurium* in pork meat. *Sci. Rep.* **7**, 1–10 (2017).
14. Voiculescu, I. & Nordin, A. N. Acoustic wave based MEMS devices for bio-sensing applications. *Biosens. Bioelectron.* **33**, 1–9 (2012).
15. Ji, J. et al. An aptamer-based shear horizontal surface acoustic wave biosensor with a CVD-grown single-layered graphene film for high-sensitivity detection of a label-free endotoxin. *Microsyst. Nanoeng.* **6**, 1–11 (2020).

16. Song, S. et al. Shear mode bulk acoustic resonator based on inclined c-Axis AlN film for monitoring of human hemostatic parameters. *Micromachines* **9**, 501 (2018).
17. Ferreira, G. N., Da-Silva, A.-C. & Tomé, B. Acoustic wave biosensors: physical models and biological applications of quartz crystal microbalance. *Trends Biotechnol.* **27**, 689–697 (2009).
18. Länge, K., Rapp, B. E. & Rapp, M. Surface acoustic wave biosensors: a review. *Anal. Bioanal. Chem.* **391**, 1509–1519 (2008).
19. Zhang, Y. & Chen, D., *Multilayer Integrated Film Bulk Acoustic Resonators* (Springer Science & Business Media, 2012).
20. Laude, V., Robert, L., Daniau, W., Khelif, A. & Ballandras, S. Surface acoustic wave trapping in a periodic array of mechanical resonators. *Appl. Phys. Lett.* **89**, 083515 (2006).
21. Gao, F. et al. Towards acoustic radiation free lamb wave resonators for high-resolution gravimetric biosensing. *IEEE Sens J* **21**, 2725–2733 (2021).
22. Hashimoto, K-Y. *Surface Acoustic Wave Devices in Telecommunications* (Springer, 2000).
23. Morgan, D. *Surface Acoustic Wave Filters: With Applications to Electronic Communications and Signal Processing* (Academic Press, 2010).
24. Vogiazi, V. et al. A comprehensive review: development of electrochemical biosensors for detection of cyanotoxins in freshwater. *ACS Sens.* **4**, 1151–1173 (2019).
25. You, C.-C. et al. Detection and identification of proteins using nanoparticle–fluorescent polymer 'chemical nose' sensors. *Nat. Nanotechnol.* **2**, 318 (2007).
26. Whitcombe, M. J. et al. The rational development of molecularly imprinted polymer-based sensors for protein detection. *Chem. Soc. Rev.* **40**, 1547–1571 (2011).
27. Eersels, K., Lieberzeit, P. & Wagner, P. A review on synthetic receptors for bioparticle detection created by surface-imprinting techniques· from principles to applications. *ACS Sens.* **1**, 1171–1187 (2016).
28. Gao, F. et al. Dual transduction on a single sensor for gas identification. *Sens. Actuators B* **278**, 21–27 (2019).
29. Gao, F., Boussaid, F., Xuan, W., Tsui, C.-Y. & Bermak, A. Dual transduction surface acoustic wave gas sensor for VOC discrimination. *IEEE Electron. Device Lett.* **39**, 1920–1923 (2018).
30. Lesson, D. A simple model of feedback oscillator noise spectrum. *Proc. IEEE* **54**, 329–330 (1966).
31. Barie, N. & Rapp, M. Covalent bound sensing layers on surface acoustic wave (SAW) biosensors. *Biosens. Bioelectron.* **16**, 979–987 (2001).